\documentclass[12pt]{article}
 \usepackage[margin=1in,footskip=0.25in]{geometry}
 \usepackage{amssymb}
\usepackage[T1]{fontenc}
\usepackage[latin9]{inputenc}
\usepackage{array}
\usepackage{booktabs}
\usepackage{mathrsfs}
\usepackage{amsmath}
\usepackage[graphicx]{realboxes}
\usepackage{esint}
\usepackage{bm}
\usepackage{hyperref}
\usepackage{mathtools}
\usepackage{lscape}
\usepackage[figuresleft]{rotating}
\usepackage{setspace}
\usepackage{subcaption}
\usepackage{parskip}
\usepackage{systeme}
\usepackage{algorithm}
\usepackage[noend]{algpseudocode}
\usepackage{float}
\usepackage[sort&compress,numbers]{natbib}
\usepackage{authblk}
\usepackage{xcolor}
\usepackage{lscape}

\normalsize
\begin{document}

\title{Agent-based modelling of sports riots}
\footnotesize
\author[$1,2$]{Alastair J. Clements}
\author[$1 \star$]{Nabil T. Fadai}
  \affil[$1$]{\footnotesize{School of Mathematical Sciences, University of Nottingham, Nottingham NG7 2RD, United Kingdom}}
   \affil[$2$]{\footnotesize{Department of Infectious Disease Epidemiology,London School of Hygiene and Tropical Medicine, London WC1E 7HT, United Kingdom}}
\affil[$\star$]{\footnotesize{Correspondence: \texttt{nabil.fadai@nottingham.ac.uk}}}

\normalsize
\maketitle

\begin{abstract}
Riots originating during, or in the aftermath of, sports events can incur significant costs in damages,  as well as large-scale panic and injuries. A mathematical description of sports riots is therefore sought to better understand their propagation and limit these physical and financial damages.  In this work, we present an agent-based modelling framework that describes the qualitative features of populations engaging in riotous behaviour. 
Agents,  pertaining to either a `rioter' or a `bystander' sub-population, move on an underlying lattice and can either be recruited or defect from their respective sub-population.  In particular, we allow these individual-level recruitment and defection processes to vary with local population density.  This agent-based modelling framework provides the unifying link between multi-population stochastic models and density-dependent reaction processes.  Furthermore, the continuum description of this ABM framework is shown to be a system of nonlinear reaction-diffusion equations and faithfully agrees with the average ABM behaviour from individual simulations. Finally,  we determine the unique correspondence between the underlying individual-level recruitment and defection mechanisms with their population-level counterparts, providing a link between local-scale effects and macroscale rioting phenomena.
\end{abstract}

\section{Introduction}

Sports riots are a worldwide phenomenon and great cause for concern, due to the financial and physical damages they can incur. Moreover, the occurrence of riots can incite a sense of fear amongst the public, with people concerned for their well-being and safety.  For example,  the riots which occurred in June 2011 in Vancouver,  upon the city home team losing the Stanley Cup ice hockey tournament,  incurred approximately C\$3.78 million in damages,  52 reported assaults, and 250 visits to emergency rooms at nearby hospitals \cite{Vancouver2011rep}.  In February 2012,  79 people were killed in a riot at a football match, when Al-Masry supporters charged the field after a victory over Al-Ahly club \cite{port-said}.   By contrast, legitimate protests associated with social reform and activism have only in rare occasions led to riotous behaviour,  as the impetus of these riots is directly linked to aggressive intervention by law enforcement officials \cite{campbell2004remote,hopkins2014football}. As such,  we specifically focus on sports riots prior to police intervention, in an effort to distinguish illegitimate riotous behaviour arising in sporting events from actions linked to peaceful protest.  While public policies have been introduced with the intention of curbing hooliganism and anti-social behaviour arising from sporting events, including football banning orders in the UK \cite{stott2006football,hopkins2014football,hester2021assessing},  many of these policies have been criticised for their impact upon civil liberties and human rights \cite{stott2006football,hopkins2014football,hester2021assessing}. 

In order to lessen and limit the negative impacts of sports riots, further understanding has  been sought from social-psychological \cite{zani, mannleonpearce, soreloser, russellpersonalities, dunning} and physiological perspectives \cite{bloodpressure,testosterone}. Theoretical studies and practical investigations have aimed to relate riot initiations and escalations to several variables, including environmental factors \cite{baron, geen, dewar}, situational factors \cite{gaskell, semyonov}, the influence of alcohol \cite{fitzpatrick, piquero, guschwan, peitersen}, and a myriad of social factors \cite{mannleonpearce, zani, arms1997, apter92, soreloser, russellpersonalities, russell,ostrowsky,caseboucher,spaaij,lewisbook,fields2007}. These studies have been conducted across a wide range of different sports, sporting events, level of play, and countries. As such, while studies investigating the relation of some factors are in agreement, others stand in conflict. Nevertheless, a common element in riotous behaviour is the rise of crowd behaviours \cite{socidtheo, idandsoctheo, granovetter}.  In particular,  \cite{granovetter} focuses on the idea of `thresholds' for an individual's participation in a group activity, suggesting that  personal thresholds can decrease as  other individuals participate. 

More recently, mathematical modelling perspectives are sought after to understand riot dynamics and implement control measures with a view to reducing consequences such as property damage (c.f.  \cite{bonnasse2018epidemiological,   London2011,  nonlinearurbancrime,  communaldisorder}).
Previous mathematical models of riots \cite{London2011}, urban crime \cite{nonlinearurbancrime}, and communal disorder \cite{communaldisorder} fit deterministic models to realistic patterns and obtained data. While these models aim to reproduce the population-level behaviour of riot dynamics, few models have been proposed that emphasise the individual-level interactions that give rise to rioting (c.f. \cite{bonnasse2018epidemiological, alsenafi2021multispecies}). One mathematical framework that is suitable for describing such individual-level interactions is by using stochastic agent-based models, whereby individuals (agents) interact with one another according to pre-defined processes on an underlying spatial grid. Such models have found great use in cell-level dynamics \cite{multi-excl,simpsoncellprolif,byrne,tissueabm}, ecology \cite{fadaipopallee}, and epidemic modelling \cite{perez2009agent,ajelli2010comparing}.

In this work, we develop a stochastic agent-based model (ABM)  that characterises individual-level mechanisms  giving rise to population-level riotous behaviour. Individual agents, classified as `rioters' or `bystanders', move on a two-dimensional square lattice restricted by exclusion processes to prevent agent overlap \cite{chowdhury,multi-excl,simpsoncellprolif, fadaiunpackallee}  and can either be recruited or defect from their respective sub-population \cite{multi-excl}.  In particular, we allow recruitment and defection processes to vary with local population density: the recruitment of bystanders changes with the number of nearby rioters, while rioters defect based on the number of nearby bystanders (c.f.  \cite{fadaiunpackallee}).  While multi-population stochastic ABMs and density-dependent reaction processes in ABMs have been previously considered separately,  the combination of these two ABM frameworks,  as we present in this work, has not been previously examined.  Consequently,  this agent-based modelling framework provides the unifying link between multi-population stochastic models and density-dependent reaction processes.  Following an examination of the qualitative features of ABM simulations,  we derive the continuum limit of the ABM in order to compare average individual-level dynamics with population-level descriptions of dynamics.  The continuum description of this ABM framework is determined to be a system of nonlinear reaction-diffusion equations that describe the migration of both sub-populations, as well as the  recruitment of bystanders and defection of rioters.  We demonstrate good agreement between the ABM and continuum descriptions, which in turn provides further understanding  of individual-level mechanisms that give rise to macroscale rioting phenomena.

\section{Results and Discussion}

In this stochastic agent-based modelling framework, we consider the population of two classes of agents, termed as `rioters'  and `bystanders', on an $X{\Delta} \times Y{\Delta}$ lattice, where ${\Delta}$ is a typical amount of space an individual occupies.  We focus on non-dimensional lattices (i.e.,  ${\Delta}=1$) and represent the location of the top right corner of each site in Cartesian co-ordinates as $(x_i,y_j)=(i,j)$, where $i=1,\dots,X$ and $j=1,\dots,Y$.  A rioter at lattice site $(i,j)$ and time $t$ is denoted as $r_{i,j}(t)$; similarly,  $b_{i,j}(t)$ represents a bystander at lattice site $(i,j)$ and time $t$. Furthermore, we employ \textit{exclusion processes} to ensure that at most one agent can occupy a lattice site at any given time \cite{chowdhury,multi-excl,simpsoncellprolif, fadaiunpackallee}.

The initial configuration of each sub-population, $r_{i,j}(0)$ and $b_{i,j}(0)$, is left to the user's choice.  If spatially uniform initial conditions are desired,  rioters and bystanders can be initially seeded on the lattice with constant probabilities $r_0$ and $b_0$.  Regardless of their initial configurations, individuals in both sub-populations move to adjacent lattice sites with unbiased direction with a single motility rate $m$.  Reflecting boundary conditions are employed on the boundaries of the lattice domain for simplicity.

The ABM also incorporates agent recruitment (a bystander becoming a rioter) and defection (a rioter becoming a bystander), where the recruitment and defection rates vary with local density \cite{fadaiunpackallee}.  As a simple metric of local density,  the recruitment and defection rates will change with how many rioters, from zero to four, are present at lattice sites in their von Neumann neighbourhoods (i.e., the adjacent North, South, East, and West lattice sites). We consider the recruitment processes to have non-negative rates \(\lambda_{r0}\), \(\lambda_{r1}\), \(\lambda_{r2}\), \(\lambda_{r3}\) and \(\lambda_{r4}\), respectively.  Similarly,  the defection process  have rates \(\lambda_{d0}\), \(\lambda_{d1}\), \(\lambda_{d2}\), \(\lambda_{d3}\) and \(\lambda_{d4}\),  due to zero, one, two, three and four neighbouring bystanders, respectively.  While the recruitment and defection rates \(\lambda_{rn}\) and \(\lambda_{dn}\) are explicitly related to local pairwise interactions of neighbours for $n\ge1$, the rates \(\lambda_{r0}\) and \(\lambda_{d0}\) can also represent \textit{global},  non-local effects of recruitment and defection processes,  including spontaneous rioting,  lack of interest that devolves into defection,  and social media influences \cite{baker2011mediated}.

Finally,  we make the additional assumption that individuals move much more often than being recruited or defecting, i.e. $m\gg \max_n(\lambda_{rn}, \lambda_{dn})$.  This assumption is a standard model simplification for fast-moving populations \cite{fadaiunpackallee}. Using a Gillespie approach \cite{gillespie1977exact}, we are able to simulate the number of both agent sub-populations as a function of time and space (Algorithm 1); a MATLAB implementation of this algorithm can be found at \url{https://github.com/nfadai/Clements2021}. 

\begin{algorithm}
\caption{Pseudocode for agent-based simulations of rioter and bystander dynamics}
\label{alg:ABM}
\begin{algorithmic}[1]
	\State Set up an \(X\times Y\) lattice and specify initial placement of rioters and bystanders;
	\State Specify counters \(Q_{r}(t)\) and \(Q_{b}(t)\);
	\State Specify recruitment rates $\lambda_{rn}$,  defection rates $\lambda_{dn}$, and motility rate $m$;
	\State Set \(t=0\) and specify terminating time $t_{\text{end}}$;
	\While{\(t<t_{\text{end}}\)}
	\State Calculate random variables \(u_{1}\) and \(u_{2}\), uniformly distributed on \([0,1]\);
	\State Select an agent at random and determine its sub-population (rioter or bystander);
	\State Compute the number of nearest neighbours \(n\) in the opposite sub-population of the chosen agent to
determine \(\lambda_{rn}\) and \(\lambda_{dn}\);
	\State Calculate propensity \(p=(m+\lambda_{dn})Q_{r}(t)+(m+\lambda_{rn})Q_{d}(t)\);
	\State Calculate time step duration \(\tau=-\ln(u_{1})/p\);
	\State \(t=t+\tau\);
	\State \(Q_{r}(t)=Q_{r}(t-\tau)\);
	\State \(Q_{b}(t)=Q_{b}(t-\tau)\);
	\If{Agent is a rioter}
	\If{\(u_{2}<m/(m+\lambda_{dn})\)}
	\State Choose a neighbouring site at random to move to;
	\If{Neighbouring site is empty}
	\State Move rioter to chosen site;
	\Else
	\State Nothing happens;
	\EndIf
	\Else
	\State Rioter becomes a bystander;
	\State \(Q_{r}(t)=Q_{r}(t)-1\);
	\State \(Q_{b}(t)=Q_{b}(t)+1\);
	\EndIf
	\Else
	\If{\(u_{2}<m/(m+\lambda_{rn})\)}
	\State Choose a neighbouring site at random to move to;
	\If{Neighbouring site is empty}
	\State Move bystander to chosen site;
	\Else
	\State Nothing happens;
	\EndIf
	\Else
	\State Bystander becomes a rioter;
	\State \(Q_{b}(t)=Q_{b}(t)-1\);
	\State \(Q_{r}(t)=Q_{r}(t)+1\);
	\EndIf
	\EndIf
	\EndWhile
\end{algorithmic}
\end{algorithm}


\subsection{ABM simulations of riots}

To examine the qualitative features of  ABM simulations, we consider various choices of recruitment and defection rates and observe the spatial and temporal evolution of the total agent population. In particular, we will focus our simulations on a particular lattice configuration that represents a single street. This geometry is obtained by using the domain \(0< x\leq 200\), \(0<y\leq 20\), which is equivalent to specifying the lattice dimensions as \(X=200\) and \(Y=20\). Furthermore,  the sub-population densities $\langle R(t) \rangle$ and  $\langle B(t) \rangle$ can be computed by averaging over multiple ABM simulations:
\begin{align}
& \langle R(t) \rangle=\frac{1}{PXY}\sum_{p=1}^{P}Q_{r,p}(t), \label{eq:ravg}
\\
&\langle B(t) \rangle=\frac{1}{PXY}\sum_{p=1}^{P}Q_{b,p}(t).\label{eq:bavg}
\end{align}
Here, $Q_{r,p}(t)$ and $Q_{b,p}(t)$ are the total number of each sub-population on the lattice at time $t$, in the $p$th identically-prepared realisation of the ABM. The total number of identically-prepared realisations is $P$; we choose $P=20$ throughout this work. Finally,  when employing spatially-dependent initial configurations that are spatially dependent in the $x$-direction alone, as will be examined in Section \ref{sec:SD},  we will also consider the sub-population densities averaged over multiple simulations and averaged in the $y$-direction alone:

\begin{align}
& \langle R(x,t) \rangle=\frac{1}{PY}\sum_{p=1}^{P}\sum_{j=1}^{Y}r_{i,j,p}(t), \label{eq:ravg2}
\\
&\langle B(x,t) \rangle=\frac{1}{PY}\sum_{p=1}^{P}\sum_{j=1}^{Y}b_{i,j,p}(t).\label{eq:bavg2}
\end{align}
Here, $r_{i,j,p}(t)$ and $b_{i,j,p}(t)$ are the rioter and bystander occupancies at lattice site $(i,j)$ at time $t$ in the $p$th identically-prepared realisation of the ABM. 

\subsection{Spatially uniform initial conditions}

We first consider results of the agent-based model for simulations beginning from spatially uniform initial conditions. We present snapshots of the two agent sub-populations for initial densities \(r_{0}=0.05\) and \(b_{0}=0.25\),  representing situations where the majority of attendees at the sports event are not inclined to riot initially.  We then consider three representative parameter sets associated with different levels of recruitment and defection:
\begin{align}
&\text{Mild Unrest:} &\quad &\lambda_{rn} =  \begin{cases} 0, n=0,1,
\\ 1, n=2,3,4,
\end{cases} 
&\quad & \lambda_{dn} \equiv    1. \label{eq:Mild}
\\
&\text{Moderate Unrest:} &\quad &\lambda_{rn} =  \begin{cases} 0, n=0,
\\ 1, n=1,2,3,4,
\end{cases} 
&\quad & \lambda_{dn} =  \begin{cases} 0, n=0,
\\ 1, n=1,2,3,4. \label{eq:Med}
\end{cases} 
\\
&\text{Severe Unrest:} &\quad &\lambda_{rn} \equiv 1,
&\quad & \lambda_{dn} =  \begin{cases} 0, n=0,1,
\\ 1, n=2,3,4.\label{eq:High}
\end{cases} 
\end{align}
In the Mild Unrest regime, rioters defect at the same rate regardless of how many bystanders are present,  while bystanders are only recruited when two or more rioters are nearby. The Severe Unrest regime swaps the recruitment and defection processes: bystanders can become rioters regardless of the number of nearby rioters, while rioters only defect when two or more bystanders are nearby. Finally, in the Moderate Unrest regime, bystanders can become rioters in the presence of at least one rioter, and vice versa for the defection processes.
For all simulations, we take  $m=100 \max_n(\lambda_{rn},\lambda_{dn})=100$ to ensure spatial uniformity is retained throughout.

\begin{figure}
\centering
\includegraphics[width=.95\textwidth]{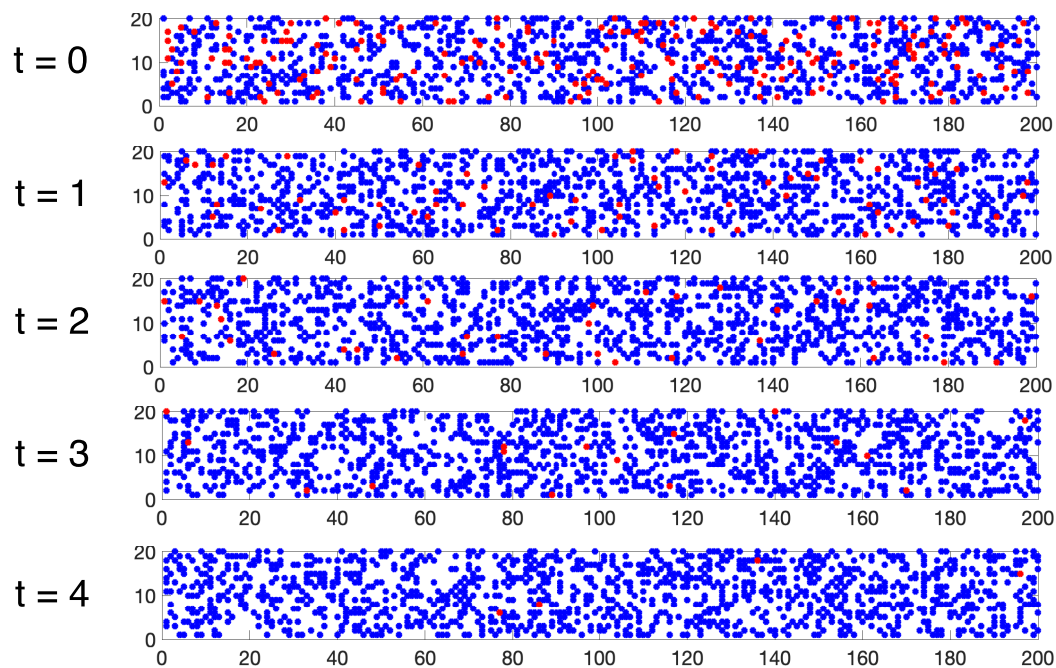}
\caption{A single realisation of rioters (red) and bystanders (blue) in the Mild Unrest parameter regime with initial densities \(r_{0}=0.05\) and \(b_{0}=0.25\).}
\label{fig:Mild1}
\end{figure}

\begin{figure}
\centering
\includegraphics[width=.95\textwidth]{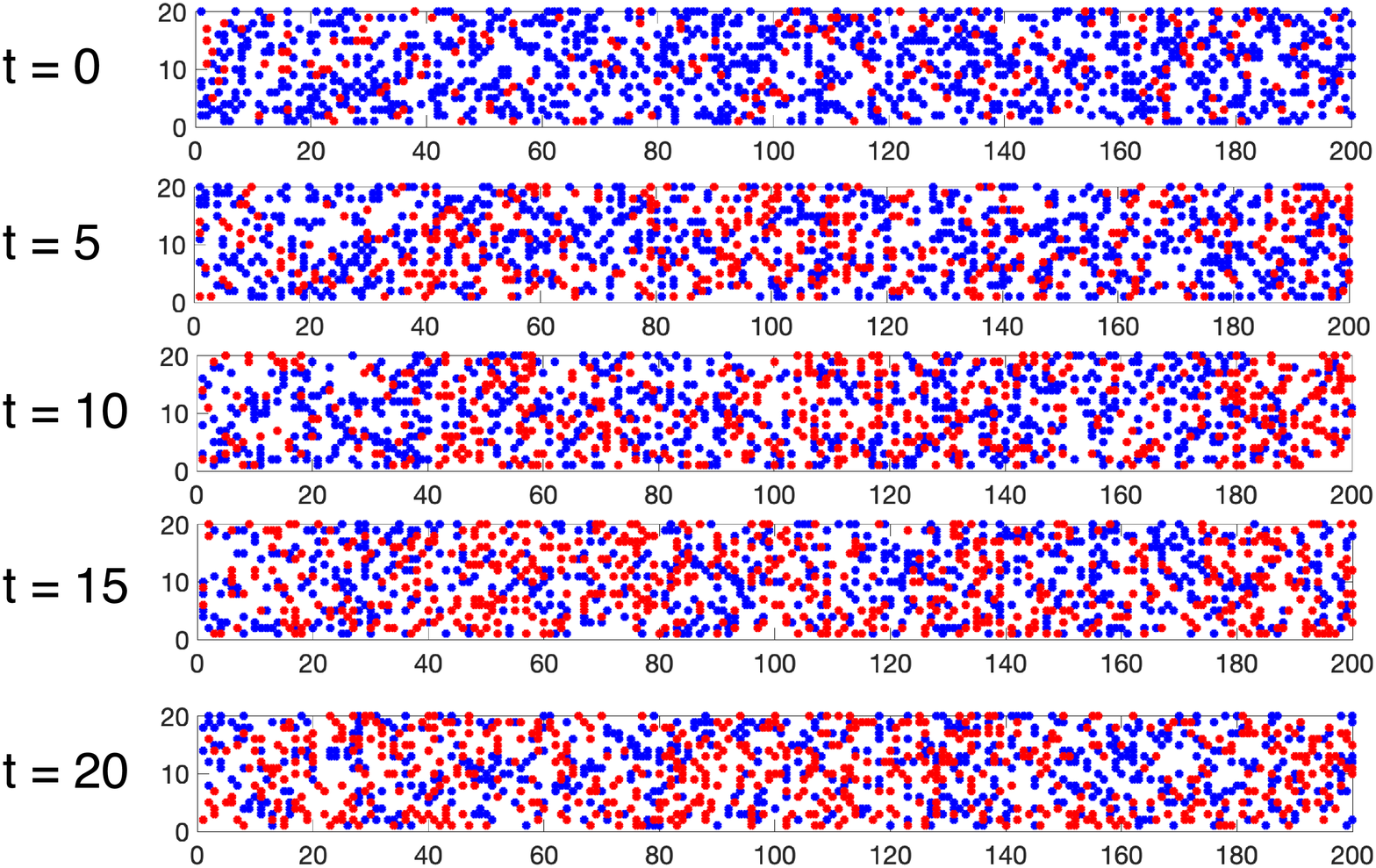}
\caption{A single realisation of rioters (red) and bystanders (blue) in the Moderate Unrest parameter regime with initial densities \(r_{0}=0.05\) and \(b_{0}=0.25\).}
\label{fig:Med1}
\end{figure}

\begin{figure}
\centering
\includegraphics[width=.95\textwidth]{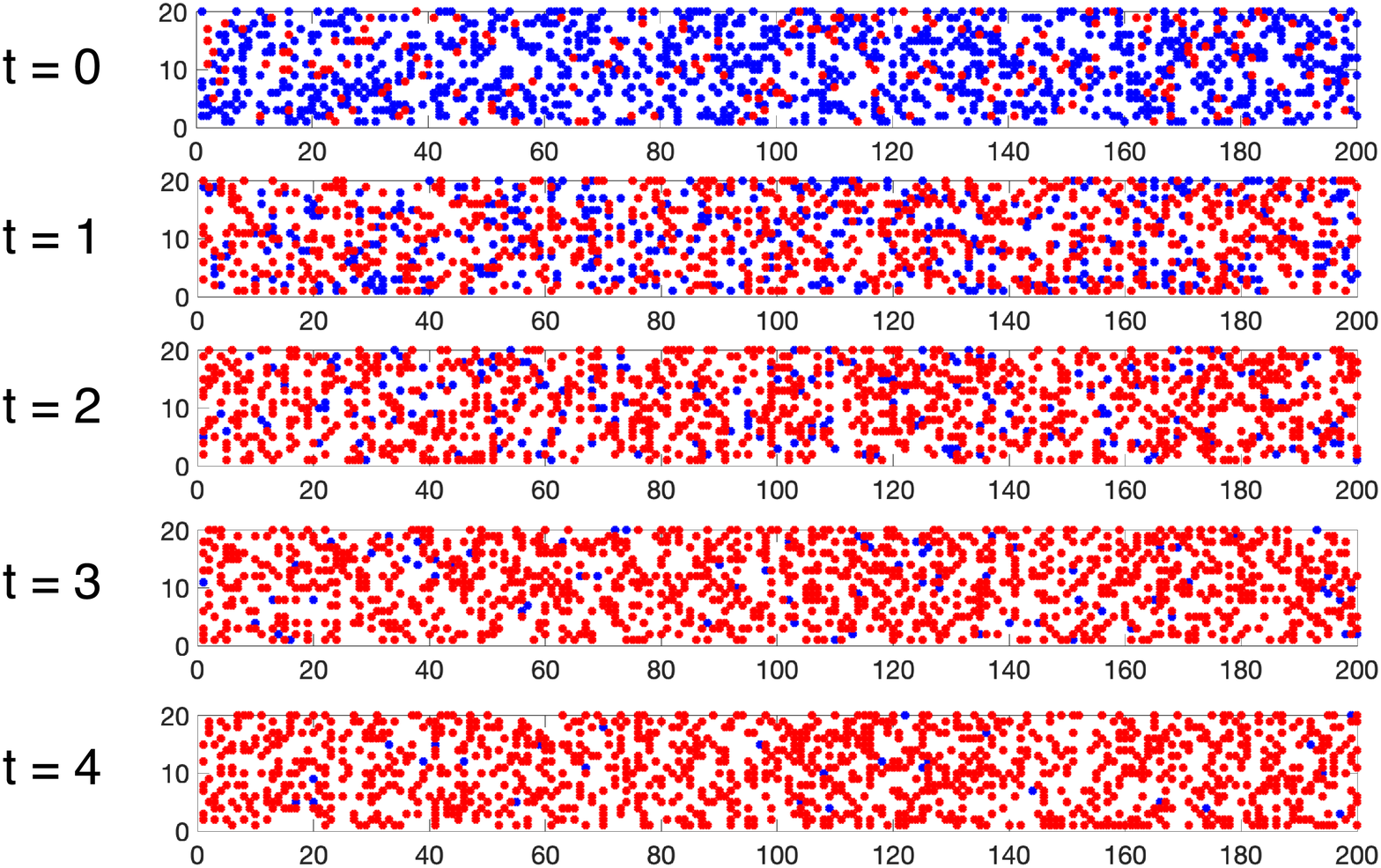}
\caption{A single realisation of rioters (red) and bystanders (blue) in the Severe Unrest parameter regime with initial densities \(r_{0}=0.05\) and \(b_{0}=0.25\).}
\label{fig:High1}
\end{figure}

Depending on the level of unrest, three main qualitative features can be observed in the agent sub-populations. In the Mild Unrest parameter regime, shown in Figure \ref{fig:Mild1}, we observe that the population eventually all become bystanders.  For larger amounts of unrest, such as the Moderate Unrest scenario shown in Figure \ref{fig:Med1}, the rioting sub-population persists, but the bystander population also persists in approximately equal numbers. Finally,  in Figure \ref{fig:High1},  we see that despite there being many more bystanders than rioters initially,  the Severe Unrest parameter regime overwhelms the defection processes and leads to the entire population becoming rioters.  While by no means a comprehensive list of phenomena,  the three unrest parameter regimes shown in Figures \ref{fig:Mild1}--\ref{fig:High1} demonstrate that the ABM framework can give rise to three main qualitative features: (i) the entire population becoming bystanders, (ii) a co-existence of rioters and bystanders,  and (iii) the entire population becoming rioters.


\subsection{Spatially uniform continuum limit}\label{sec:CL}

While the ABM framework allows us to visualise individual simulations of rioting dynamics,  it is often more convenient to examine a simpler mathematical description of the average behaviour of the ABM, called the \textit{continuum limit description} \cite{compart-based, multi-excl,  fadaiunpackallee}.  The continuum limit description gives us the ability to study global,  deterministic features of the ABM when the number of lattice sites is large and the number of simulations being averaged is also large.  As a result, we can compare the average ABM sub-population densities, $\langle R(t) \rangle$ and  $\langle B(t) \rangle$,  with their continuum limit analogues, denoted as $r(t)$ and $b(t)$ respectively.

When the ABM employs spatially uniform initial conditions and the motility rate of agents $m$ is large,  the net flux of agents entering and leaving each lattice site due to motility events is,  on average,  zero \cite{fadaiunpackallee}. Therefore,  spatial derivatives in the continuum limit will vanish,  meaning that the continuum description of the average sub-population densities,  $0\le r,  b \le 1$,  are functions of time alone.  For the derivation of the continuum limit of each sub-population, we follow \cite{compart-based,  fadaiunpackallee} and consider each recruitment and defection processes individually.  For recruitment of bystanders to rioters at rate \(\lambda_{rn}\),  we need to consider all the spatial configurations for which a bystander has precisely \(n\) neighbouring sites occupied by rioters, and precisely \(4-n\) sites not occupied by rioters.  Similarly,  for the defection of rioters to bystanders at rate \(\lambda_{dn}\),  a rioter must have exactly \(n\) neighbouring sites occupied by bystanders and the remaining \(4-n\) sites  not occupied by bystanders.  Accounting for all of these possibilities leads to the following continuum limit descriptions for \(r(t)\) and \(b(t)\):
\begin{equation}
	\frac{\mathrm{d}r}{\mathrm{d}t}=-\frac{\mathrm{d}b}{\mathrm{d}t}=\underbrace{b\sum_{n=0}^{4}{\lambda_{rn}{4\choose{n}}r^{n}(1-r)^{4-n}}}_{\text{recruitment}}-\underbrace{r\sum_{n=0}^{4}{\lambda_{dn}{4\choose{n}}b^{n}(1-b)^{4-n}}}_{\text{defection}}. \label{eq:CL}
\end{equation}

Furthermore,  due to the ABM reflecting boundary conditions and lack of any source or sink terms in the ABM framework, the total number of agents is conserved:
\begin{equation}
r(t)+b(t)=r_0+b_0:=K\le1.
\end{equation}
Therefore, we can rearrange \eqref{eq:CL} in terms of $r(t)$ alone:

\begin{equation}
	b(t)=K-r(t), \qquad \frac{\mathrm{d}r}{\mathrm{d}t}=(K-r)\sum_{n=0}^{4}{\lambda_{rn}{4\choose{n}}r^{n}(1-r)^{4-n}}-r\sum_{n=0}^{4}{\lambda_{dn}{4\choose{n}}(K-r)^{n}(1-K+r)^{4-n}}. \label{eq:CL2}
\end{equation}

\subsubsection{Comparison of ABM agent density and continuum limit}
To highlight the similarities between the continuum limit and the average behaviour of ABM simulations, we examine the population density of each sub-population in the parameter regimes described in equations \eqref{eq:Mild}--\eqref{eq:High}.  From \eqref{eq:CL2},  the corresponding continuum limit descriptions of the rioter density for each parameter regime become the following:
\begin{align}
&\text{Mild Unrest:} &\quad&\frac{\mathrm{d}r}{\mathrm{d}t}= (K-r)r^2(3r^2-8r+6)-r, \label{eq:CLmild}
\\
&\text{Moderate Unrest:} &\quad&\frac{\mathrm{d}r}{\mathrm{d}t}= (K-r)[1-(1-r)^4]-r[1-(1-K+r)^4],\label{eq:CLmed}
\\
&\text{Severe Unrest:} &\quad&\frac{\mathrm{d}r}{\mathrm{d}t}= (K-r)-r(K-r)^2[3(K-r)^2-8(K-r)+6].\label{eq:CLhigh}
\end{align}

In the Mild Unrest case,  the only steady-state for $r,b\in[0,K]$ is $(r,b)=(0,K)$,  which is stable.  Similarly, the Severe Unrest case only has $(r,b)=(K,0)$ as a steady-state,  which is stable. Finally, in the Moderate Unrest case, there are three steady-states: $(r,b)=(0,K),(K/2,K/2),(K,0)$, which are unstable, stable, and unstable, respectively. While only a small representative of the sample parameter space, the continuum limit equations for $r$ and $b$ clearly show  the possibility of three steady-state values for $r$: no rioters ($r=0$), all rioters ($r=K$) and an intermediate rioter population density in the interval $(0,K)$.  

\begin{figure}
\centering
\includegraphics[width=0.45\textwidth]{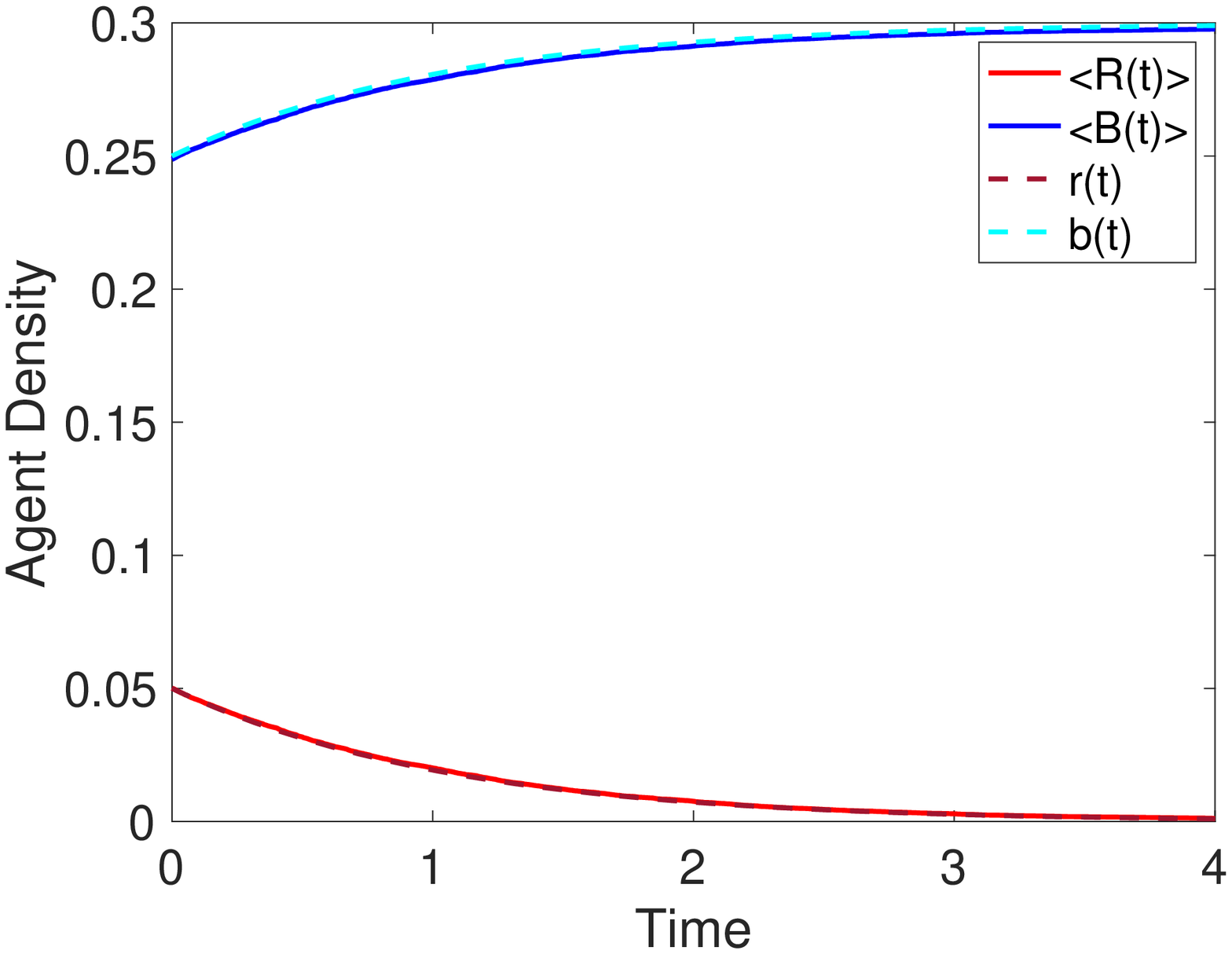}
\includegraphics[width=0.45\textwidth]{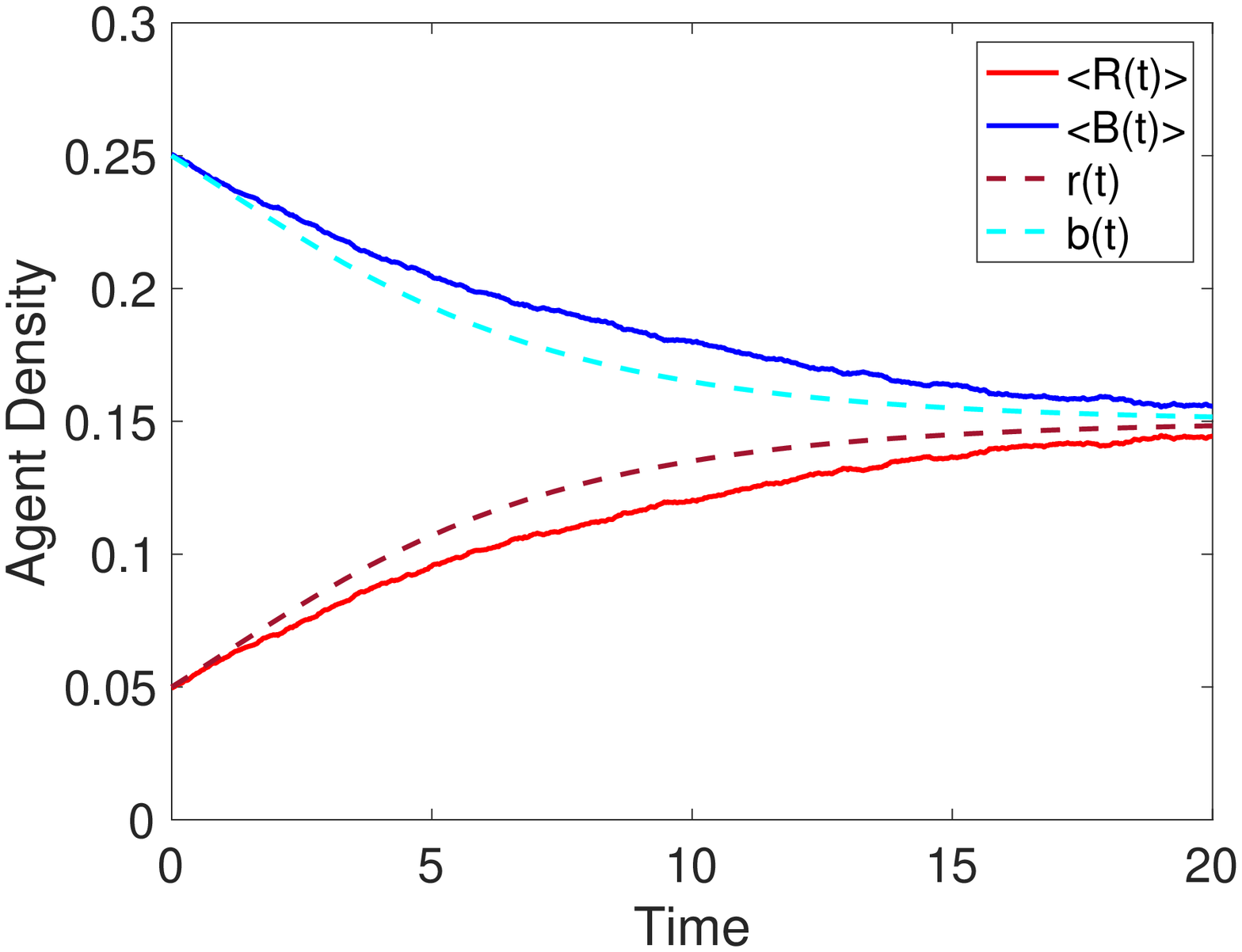}
\\
(a) \hspace{7cm} (b) 
\\
\includegraphics[width=0.45\textwidth]{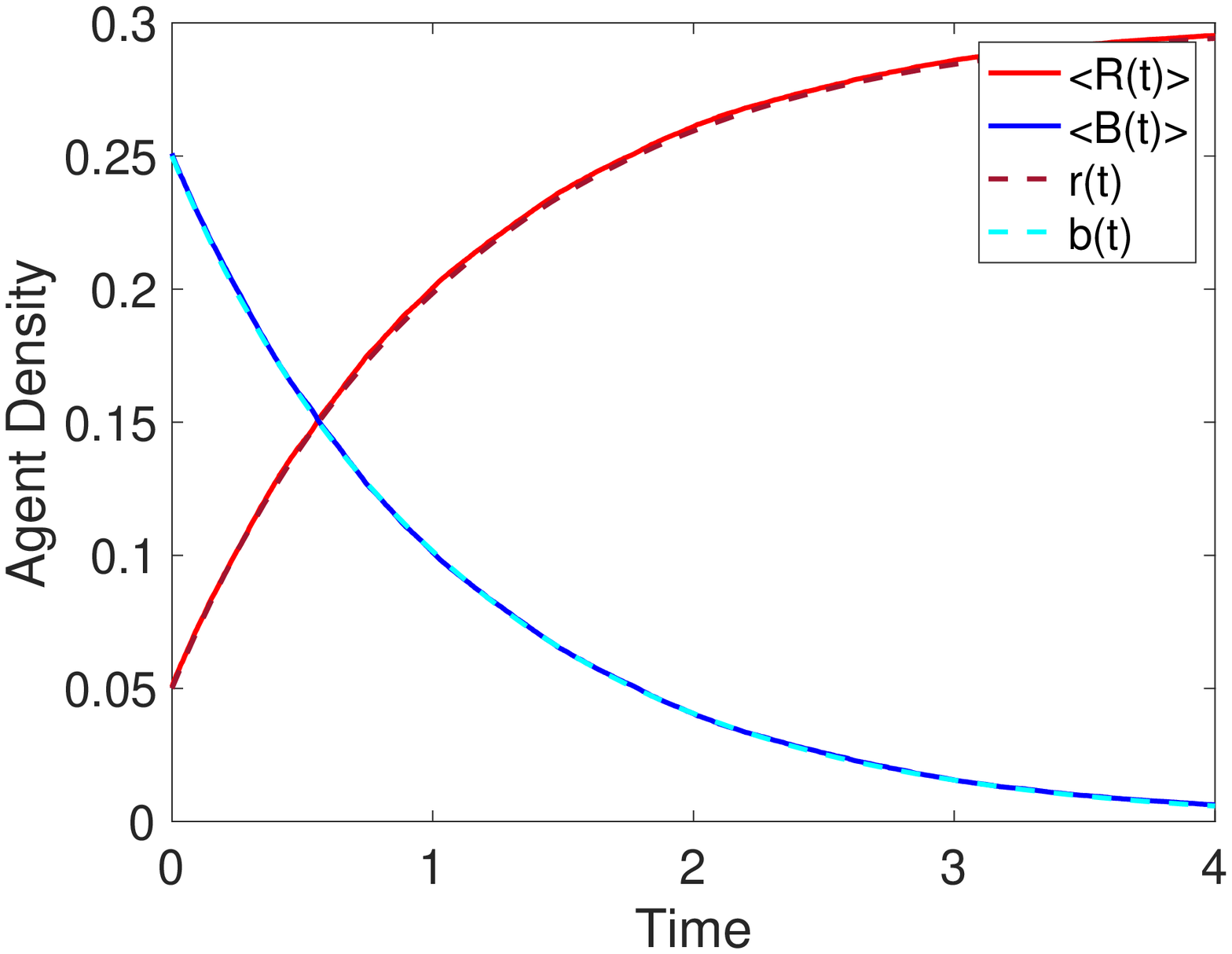}
\\
(c)
\caption{Comparison of the average ABM behaviour over 20 identically-prepared simulations, $\langle R(t) \rangle$ and $\langle B(t) \rangle$,  with their continuum limit descriptions, $r(t)$ and $b(t)$.  All simulations begin with the initial densities \(r_{0}=0.05\) and \(b_{0}=0.25\) and the parameter regimes used are: (a) Mild Unrest; (b) Moderate Unrest; and (c) Severe Unrest.}
\label{fig:CLvsABM}
\end{figure}

In Figure \ref{fig:CLvsABM}, we compare average ABM behaviour over 20 identically-prepared simulations, $\langle R(t) \rangle$ and $\langle B(t) \rangle$ defined in \eqref{eq:ravg} and \eqref{eq:bavg} and $P=20$,  with their continuum limit descriptions, $r(t)$ and $b(t)$ defined in  \eqref{eq:CL2}.  The numerical solutions of \eqref{eq:CL2} are computed using \texttt{ode45} in MATLAB.  We observe excellent agreement between the ABM and continuum descriptions of agent densities in the Mild and Severe Unrest regimes.  In the Moderate Unrest regime, we note that while the same equilibrium density value is achieved, there is some discrepancy between the two model descriptions for intermediate time.  As some continuum limit descriptions of ABM frameworks require additional refinements for accuracy, include agent state space, agent adhesion, and clustering effects (c.f. \cite{compart-based, gapfilling,johnston2020predicting,fadaiunpackallee}),  we anticipate that the Moderate Unrest parameter regime will require additional terms in the continuum limit description for more accuracy.

\subsection{Determining individual-level mechanisms from global population dynamics: inverse problem}
It is important to emphasise at this point that the three parameter regimes considered in this section (Mild, Moderate and Severe Unrest) are by no means an exhaustive list of potential phenomena that can occur as predicted via the continuum limit. Since \eqref{eq:CL2} reduces to a polynomial in $r$ of degree 5, it is possible to have up to 5 equilibria in $[0,K]$.  Additionally, it is more likely that we will know the \textit{global} trends in agent and bystander populations rather than their \textit{local}, individual-based mechanisms of rioting or defecting. Consequently,  we will now explore the \textit{inverse problem} of obtaining the local recruitment and defection rates, i.e. $\lambda_{rn}$ and $\lambda_{dn}$, from a given continuum description of a particular rioter sub-population.

To solve this inverse problem, we follow \cite{fadaiunpackallee} and apply the same methodologies to relate the continuum limit of a particular ABM parameter set to a given global population description of rioters.  Firstly, we rewrite the continuum limit system shown in \eqref{eq:CL} in terms of Bernstein basis polynomials of fourth degree \cite{bernstein}:

\begin{equation}
	\frac{\mathrm{d}r}{\mathrm{d}t}=-\frac{\mathrm{d}b}{\mathrm{d}t}=b\sum_{n=0}^{4}\lambda_{rn}B_{n,4}(r) -r\sum_{n=0}^{4}\lambda_{dn}B_{n,4}(b),
\end{equation}
where 
\begin{equation}
B_{n,4}(x)={4\choose{n}}x^{n}(1-x)^{4-n},\quad n=0,1,2,3,4.
\end{equation}
We can then convert these Bernstein basis functions to the standard basis of monomials \(\{x^{0},x^{1},x^{2},x^{3},x^{4}\}\), by means of the following transformation \cite{farouki1987}:
\begin{equation}
	x^{m}=\sum_{n=m}^{4}{\frac{{n\choose{m}}}{{4\choose{m}}}B_{n,4}(x)}
\iff	\mathbf{x}=\mathbf{M}\mathbf{b},	
\end{equation}
where

\begin{align}
	\mathbf{x}=
	\begin{bmatrix}
		x^{0}\\
		x^{1}\\
		x^{2}\\
		x^{3}\\
		x^{4}\\
	\end{bmatrix},
	&& \mathbf{M}=
	\begin{bmatrix}
		1 & 1 & 1 & 1 & 1\\
		0 & 1/4 & 1/2 & 3/4 & 1\\
		0 & 0 & 1/6 & 1/2 & 1\\
		0 & 0 & 0 & 1/4 & 1\\
		0 & 0 & 0 & 0 & 1
	\end{bmatrix},
	&& \mathbf{b}=
	\begin{bmatrix}
		B_{0,4}(x)\\
		B_{1,4}(x)\\
		B_{2,4}(x)\\
		B_{3,4}(x)\\
		B_{4,4}(x)\\
	\end{bmatrix}.\label{matrixeq}
\end{align}

This one-to-one transformation enables us to directly identify population-level parameters with corresponding individual rates.  In other words,  if we assume that the population-level descriptions of recruitment and defection processes are expressed as

\begin{equation}
	\frac{\mathrm{d}r}{\mathrm{d}t}=-\frac{\mathrm{d}b}{\mathrm{d}t}=b\sum_{n=0}^{4}\alpha_n r^n -r\sum_{n=0}^{4}\delta_n b^n,
\end{equation}
we are able to identify, by means of the Bernstein basis transformation, that
\begin{equation}
\sum_{n=0}^{4}\alpha_n r^n =  \sum_{n=0}^{4}B_{n,4}(r) \left[\alpha_0 + \frac{\alpha_1 n}{4}+\frac{\alpha_2 n(n-1)}{12} + \frac{\alpha_3 n(n-1)(n-2)}{4!}+\frac{\alpha_4 n(n-1)(n-2)(n-3)}{4!} \right],
\end{equation}
which immediately implies that
\begin{align}
\lambda_{r0}&=\alpha_0,
\\
\lambda_{r1}&=\alpha_0+\frac{\alpha_1}{4},
\\
\lambda_{r2}&=\alpha_0+\frac{\alpha_1}{2}+\frac{\alpha_2}{6},
\\
\lambda_{r3}&=\alpha_0+\frac{3\alpha_1}{4}+\frac{\alpha_2}{2}+\frac{\alpha_3}{4},
\\
\lambda_{r4}&=\alpha_0+\alpha_1+\alpha_2+\alpha_3+\alpha_4.
\end{align}
A near-identical calculation can be used to relate the global defection rate parameters, $\delta_n$, with their corresponding individual-level parameters,  $\lambda_{dn}$. For ease of computation, it is worth noting that the individual-level rates $\lambda_{rn}$ can also be obtained by multiplying each row of $\mathbf{M}$ in \eqref{matrixeq} by their corresponding $\alpha_m$ values and summing the $n$th column.

\subsubsection{A caveat on individual-level parameter identifiability}
At this point,  we should stress that the identifiability of these individual-level recruitment and defection mechanisms can only be uniquely determined if the global recruitment and defection rates are known separately to one another. Contrastingly, if only the \textit{net} global sub-population growth rate is known, the majority of the individual-level rates cannot be uniquely determined.  To demonstrate this claim, suppose that the net sub-population growth of rioters is known to be a polynomial of degree 5 of fewer:

\begin{equation}
 \frac{\mathrm{d}r}{\mathrm{d}t}=G(r) := \sum_{m=0}^{5} \beta_m r^m. \label{eq:Inv0}
\end{equation} 
As the continuum limit shown in \eqref{eq:CL2}, i.e., the rioter sub-population growth rate,  is also a polynomial of degree 5 or fewer,  we can attempt to determine unique choices of $\lambda_{rn}$ and $\lambda_{dn}$ that will identically match $G(r)$:

\begin{equation}
 \frac{\mathrm{d}r}{\mathrm{d}t}=(K-r)\sum_{n=0}^{4}{\lambda_{rn}{4\choose{n}}r^{n}(1-r)^{4-n}}-r\sum_{n=0}^{4}{\lambda_{dn}{4\choose{n}}(K-r)^{n}(1-K+r)^{4-n}} = \sum_{m=0}^{5} \beta_m r^m. \label{eq:Inv}
\end{equation} 

It immediately follows that, due to 10 unknown parameters on the left hand side of \eqref{eq:Inv} being matched to 6 known parameters on the right hand side of \eqref{eq:Inv}, the associated inverse problem is underdetermined.  However, by evaluating \eqref{eq:Inv} at $r=0,K$, we are able to uniquely determine two of the individual-level rates, $\lambda_{r0}$ and $\lambda_{d0}$:
\begin{equation}
\lambda_{r0}=\frac{\beta_0}{K},\qquad \lambda_{d0}=-\sum_{m=0}^{5} \beta_m K^{m-1}.
\end{equation}
Since all individual-level rates are assumed to be non-negative, it follows that two key constraints of the global recruitment rate are
\begin{equation}
\beta_0\ge0,\qquad \sum_{m=0}^{5} \beta_m K^m \le0.
\end{equation}
In other words, the recruitment rate at $r=0$ must be non-decreasing, while the recruitment rate at $r=K$ must be non-increasing; both of these constraints are expected since the total number of agents must remain constant \cite{fadaiunpackallee}.

The remaining eight individual-level recruitment and defection rates can be related by equating powers of $r^m, $ for $m=1,2,...,5$. However, we will still have at least three degrees of freedom in this reduced underdetermined system. As an illustrative example of the non-identifiability of the individual-level rates, let us consider a rioter growth rate that behaves akin to logistic growth (c.f.  \cite{fadaipopallee,murray,fadaiunpackallee}):
\begin{equation}
 \frac{\mathrm{d}r}{\mathrm{d}t}=r(K-r). \label{eq:Inv2}
\end{equation} 
It can be shown that there are four freely chosen parameters, $\lbrace A,B,C,D\rbrace$, that emerge when decomposing this rioter growth rate into a difference of recruitment and defection rates:
\begin{equation}
 \frac{\mathrm{d}r}{\mathrm{d}t}=(K-r)[(1+A)r+Br^2+Cr^3+Dr^4]-r
 [A(K-r)+Br(K-r)+Cr^2(K-r)+Dr^3(K-r)].
 \end{equation}
Furthermore,  by using the aforementioned Bernstein basis transformation shown in \eqref{matrixeq}, we determine that the individual-level recruitment and defection rates are
\begin{align*}
\lambda_{r0}&=\lambda_{d0}=0,
\\
\lambda_{r1}&=\frac{1+A}{4},
\\
\lambda_{r2}&=\frac{1+A}{2}+\frac{	B}{6},
\\
\lambda_{r3}&=\frac{3(1+A)}{4}+\frac{B}{2}+\frac{C}{4},
\\
\lambda_{r4}&=1+A+B+C+D,
\\
\lambda_{d1}&=\frac{A+KB+K^2C+K^3D}{4},
\\
\lambda_{d2}&=\frac{A+KB+K^2C+K^3D}{2}-\frac{B+2KC+3K^2D}{6},
\\
\lambda_{d3}&=\frac{3A}{4}+\frac{B(3K-2)}{4}-\frac{C(1-K)(1-3K)}{4}+\frac{3K(1-K)^2D}{4},
\\
\lambda_{d4}&=A-(1-K)B+(1-K)^2C-(1-K)^3D.
\end{align*}
While we require that all of these individual-level rates are non-negative,  there is still a considerable subspace within  $\lbrace A,B,C,D\rbrace$-space to pick different individual-level rates that give rise to the same rioter growth rate.

To summarise, the key features of the ABM while employing spatially uniform initial conditions give rise to three main qualitative features: complete take-over by rioters, complete take-over of bystanders, or a co-existence equilibrium of both sub-populations. All three qualitative features are faithfully reproduced in the continuum limit of the ABM, which also gives rise to a systematic method of relating individual-level recruitment and defection rates to their analogous population-level counterparts. However, these individual-level rates cannot be uniquely determined if only the \textit{net} growth mechanisms of either sub-population,  i.e. the net difference between recruitment and defection rates, is known. Nevertheless, the associated individual-level mechanisms can be obtained with the inclusion of a few freely-determined parameters.

\subsection{Spatially-dependent initial conditions}\label{sec:SD}

\begin{figure}
\centering
\includegraphics[width=.95\textwidth]{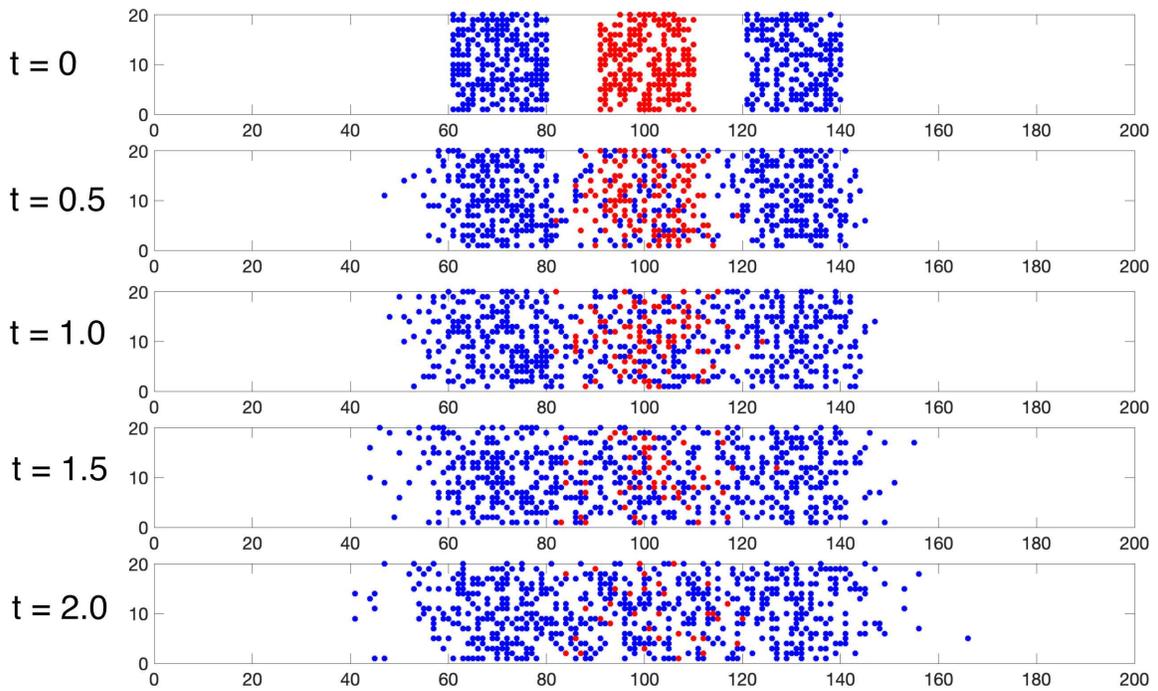}
\caption{A single realisation of rioters (red) and bystanders (blue) in the Mild Unrest parameter  regime with initial conditions listed in  \eqref{eq:PDE_ICr}-- \eqref{eq:PDE_ICb}.}
\label{fig:Mild2}
\end{figure}

\begin{figure}
\centering
\includegraphics[width=.95\textwidth]{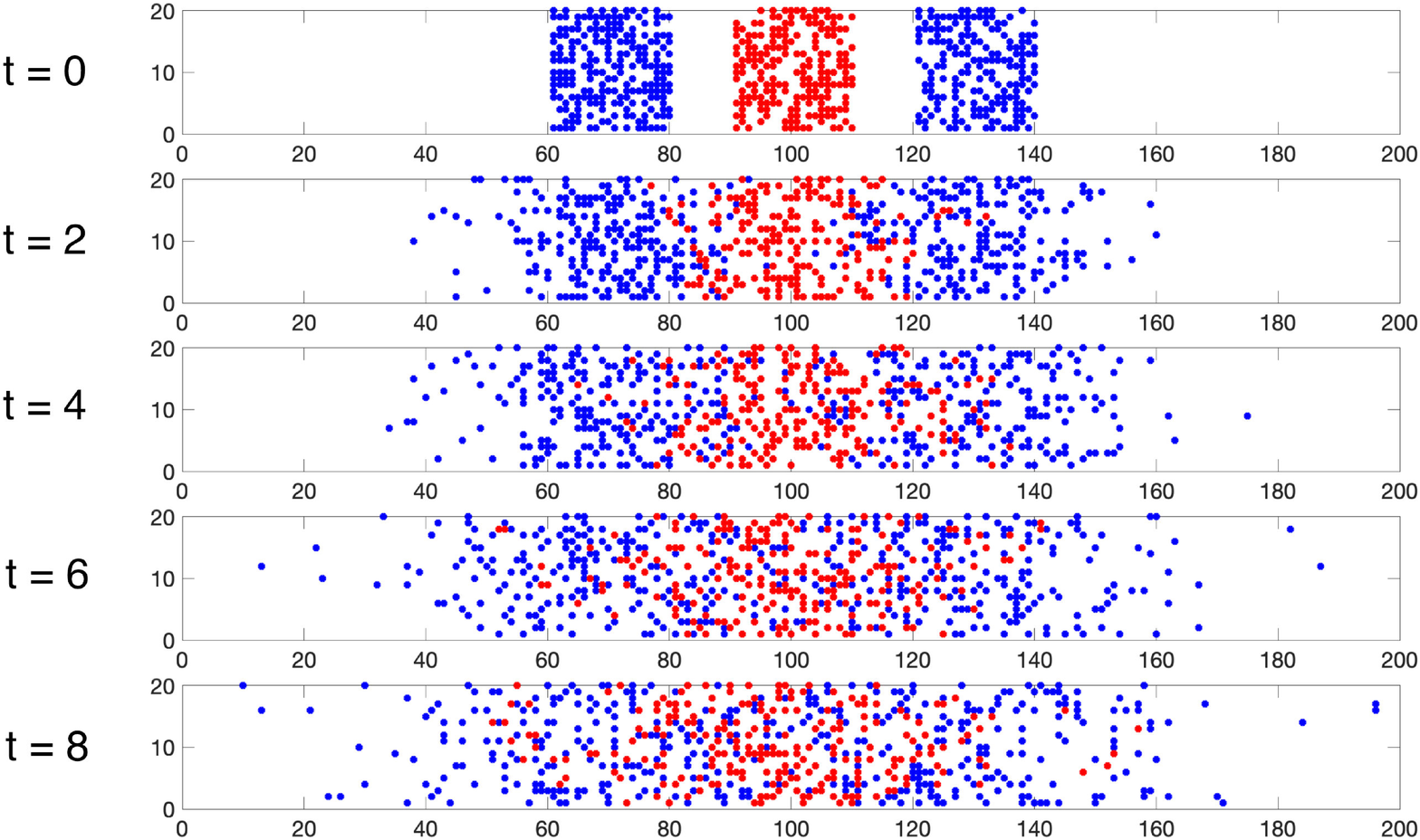}
\caption{A single realisation of rioters (red) and bystanders (blue) in the Moderate Unrest parameter regime with initial conditions listed in  \eqref{eq:PDE_ICr}-- \eqref{eq:PDE_ICb}.}
\label{fig:Med2}
\end{figure}

\begin{figure}
\centering
\includegraphics[width=.95\textwidth]{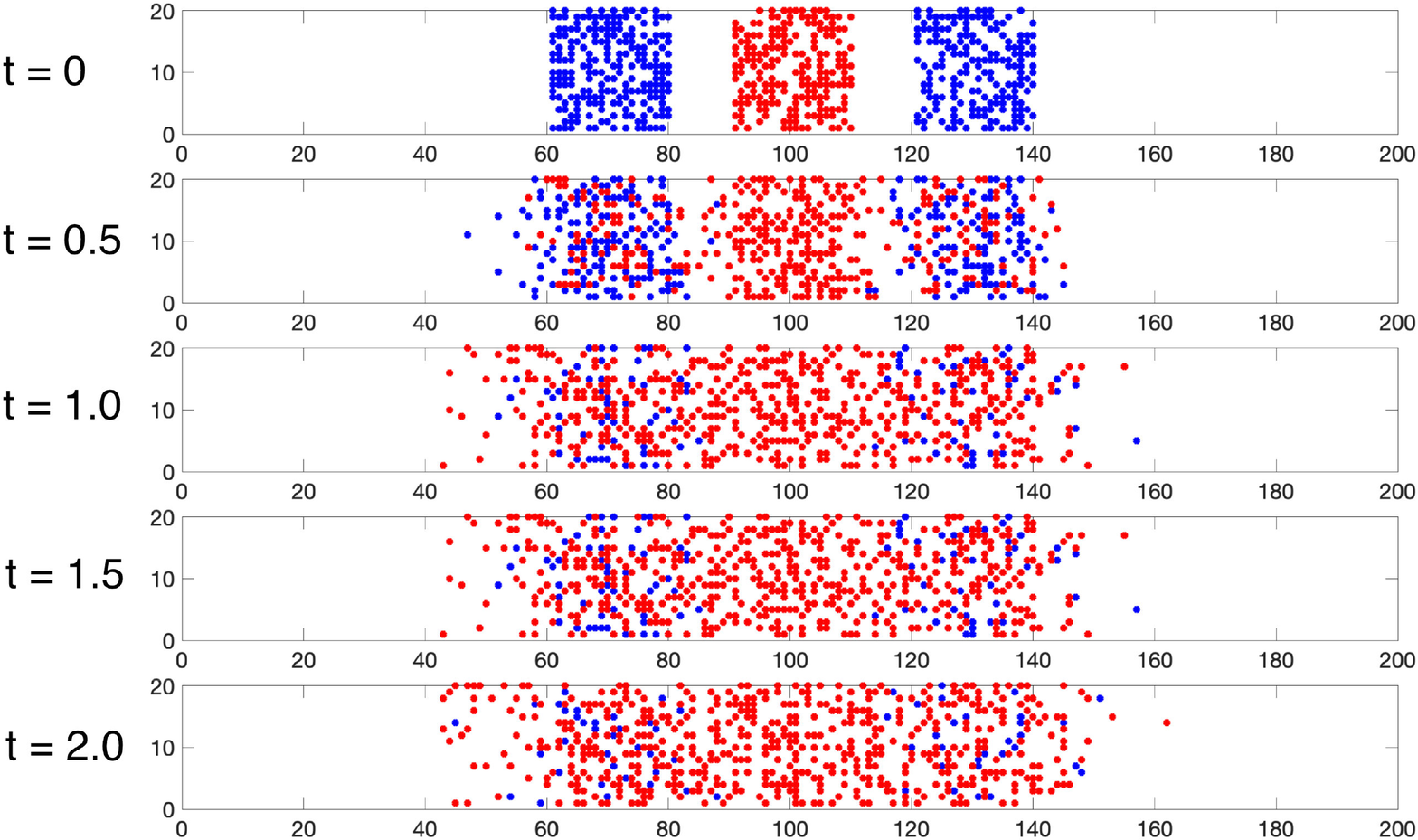}
\caption{A single realisation of rioters (red) and bystanders (blue) in the Severe Unrest parameter regime with initial conditions listed in  \eqref{eq:PDE_ICr}-- \eqref{eq:PDE_ICb}.}
\label{fig:High2}
\end{figure}

To incorporate spatial dependence within ABM simulations, we can employ spatially-dependent initial conditions in the ABM framework to observe how sub-population densities evolve in both space and time. This is analogous to considering situations whereby supporters of a particular sports team are grouped together and become riotous upon their team losing the game. For this spatial configuration, we consider a `block' of rioters with average population density $r_0$ centred along the street, while blocks of bystanders  with average population density $b_0$  are initially on either side of the rioters:

\begin{align}
    r_{i,j}(0)&= \begin{cases}
              r_0, &  91\leq i\leq 110, 1\leq j\leq 20,\\
               0, & \text{otherwise.}
          \end{cases}\label{eq:PDE_ICr}\\
    b_{i,j}(0)&= \begin{cases}
              b_0, &  61\leq i\leq 80 \text{ or } 121\leq i\leq 140, 1\leq j\leq 20,\\
               0, & \text{otherwise.}
          \end{cases}\label{eq:PDE_ICb}
\end{align}
While the initial population densities $r_0, b_0$ can be set to 1,  as is often chosen with spatially-dependent ABM simulations (c.f. \cite{fadaiunpackallee, multi-excl, compart-based}),  we will assign the initial population densities $r_0=b_0=0.5$ for simulations shown in  Figures \ref{fig:Mild2}--\ref{fig:High2}. This reduced initial population density is to prevent any local clustering from hindering recruitment or defection processes at the individual scale.  Furthermore, it is unrealistic that groups of people will be packed as close as physically possible in a block, whereas cells as other populations previously considered in similar ABM simulations can easily achieve maximum population density in a given region (c.f. \cite{multi-excl, compart-based}).


\begin{figure}
\centering
\includegraphics[width=\textwidth]{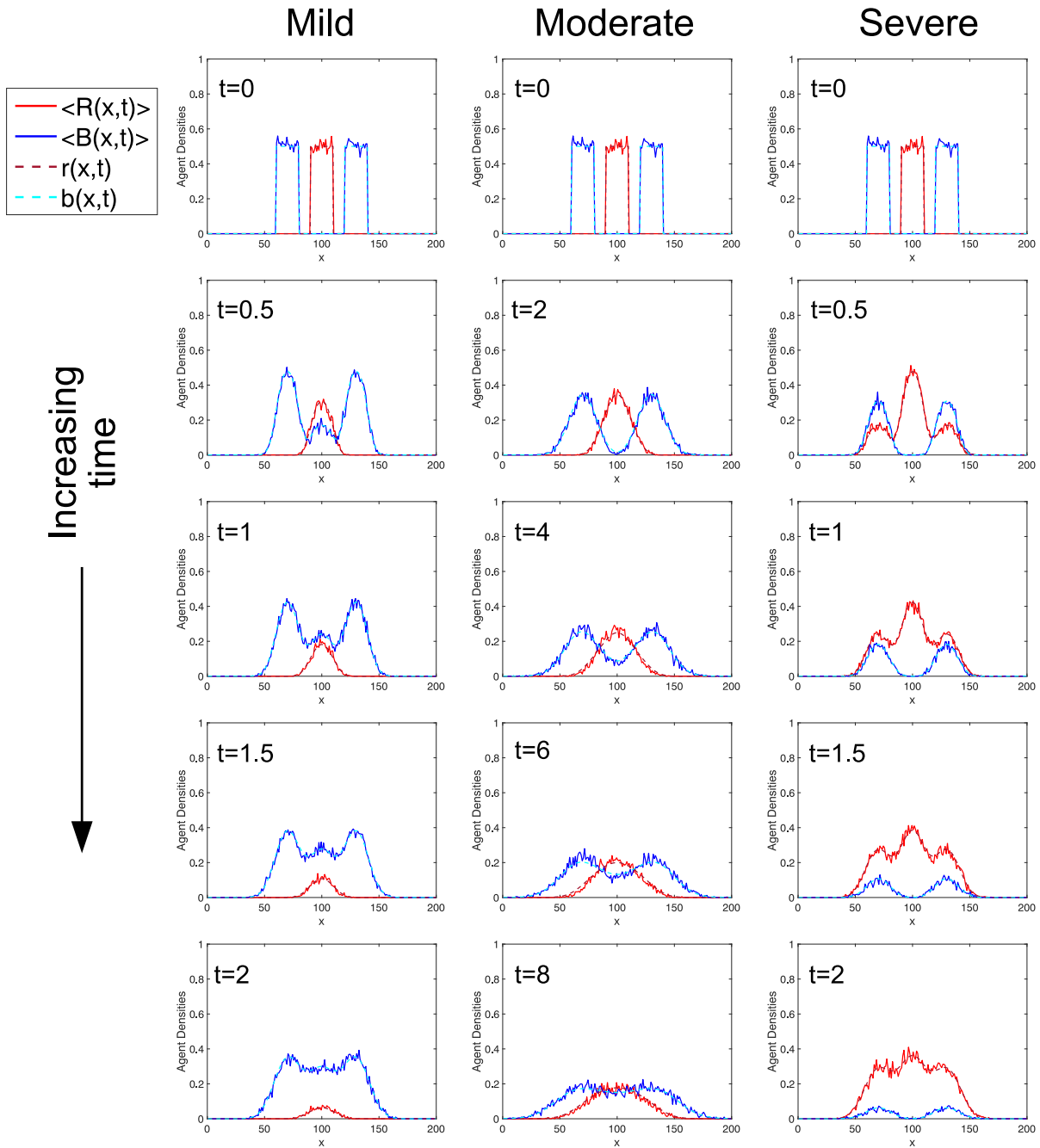}
\caption{Comparison of the average ABM behaviour over 20 identically-prepared simulations, $\langle R(x,t) \rangle$ and $\langle B(x,t) \rangle$,  with their continuum limit descriptions, $r(x,t)$ and $b(x,t)$.  All simulations begin with the initial conditions described in \eqref{eq:PDE_ICr}--\eqref{eq:PDE_ICb} and the three parameter regimes (Mild Unrest,  Moderate Unrest,  and Severe Unrest) are described in \eqref{eq:Mild}--\eqref{eq:High}.}
\label{fig:PDEvsCL}
\end{figure}

To modify the continuum limit of the ABM to incorporate spatial dependence, we follow \cite{multi-excl} to determine the effects of diffusion and motility within the continuum limit. Combined with the aforementioned recruitment and defection processes stated in Section \ref{sec:CL}, we have that the continuum limit description of the ABM is represented as a coupled PDE system for  \(r(x,y,t)\) and \(b(x,y,t)\):
\begin{align}
	\frac{\partial r}{\partial t}&=D\nabla \cdot \left[(1-b)\nabla r+r\nabla b\right]+\rho(r,b), \label{continuumr}
	\\
	\frac{\partial b}{\partial t}&=D\nabla \cdot \left[(1-r)\nabla b+b\nabla r\right]-\rho(r,b),  \label{continuumb}
\end{align}
where
\begin{equation}
D=\frac{m\Delta^{2}}{4} ~~\text{ and }~~ \rho(r,b) =b\sum_{n=0}^{4}\lambda_{rn}B_{n,4}(r) -r\sum_{n=0}^{4}\lambda_{dn}B_{n,4}(b).  \label{Dsum}
\end{equation}
We note that, due to the reflecting boundary conditions and the initial conditions being independent of $y$, the solutions for $r$ and $b$ will also be independent of $y$ \cite{pathlines}, i.e.  $r(x,y,t)=r(x,t)$ and $b(x,y,t)=b(x,t)$. Additionally, the incorporation of linear and cross-diffusion terms in the continuum limit descriptions do not affect the underlying recruitment and defection rates discussed previously. In other words, the Mild, Moderate, and Severe parameter regimes described in \eqref{eq:Mild}--\eqref{eq:High} continue to obey the continuum limit descriptions shown in \eqref{eq:CLmild}--\eqref{eq:CLhigh}.
Furthermore,  by combining \eqref{continuumr} with \eqref{continuumb}, we note that the total number of agents, $T=r+b$,  continue to be a conserved quantity within the domain, while the evolution of agents within the domain follows the standard linear diffusion equation:
\begin{equation}
\frac{\partial T}{\partial t}=D\nabla^2 T.
\end{equation}
Finally,  as discussed in  \cite{multi-excl},  each sub-population density evolves according to standard linear diffusion when a single sub-population is present, whereas cross-diffusion effects play a larger role when both sub-populations are present. Numerical solutions of the PDE system \eqref{continuumr}--\eqref{Dsum}, such as those presented in Figure \ref{fig:CLvsABM},  are computed using \texttt{pdepe} in MATLAB.

With reference to Figure \ref{fig:PDEvsCL}, we observe that the continuum limit of the ABM faithfully reproduces the average behaviour of ABM simulations employing spatially-dependent initial conditions. Like in the case where spatially uniform initial conditions are employed, the Mild and Severe Unrest parameter regimes evolve over faster timescales than the Moderate Unrest parameter regime, since agents in the Mild and Severe Unrest parameter regimes can undergo spontaneous defection or recruitment without the requirement of  agents from the opposing sub-population to be present.

\section{Conclusions}

In this work, we propose a new agent-based model (ABM) that can be used to simulate individuals involved in sports riots. Unlike other forms of rioting, which are often escalated and exacerbated due to the presence of law enforcement officials,  sports riots are generally initiated from within a sub-population of sports-goers. With a view to limit property damage and contain anti-social behaviour resulting from sports riots, it is essential to understand the temporal and spatial evolution of the aforementioned rioting sub-population. 

To provide a qualitative understanding of the rioting phenomena that can arise from simulations of sports riots, we consider an ABM with two sub-populations (rioters and bystanders), in which agents can move and change sub-population type by means of recruitment and defection mechanisms. These individual-level mechanisms vary with the local population density of the opposite sub-population and can be shown to be linked in one-to-one correspondence with prescribed \textit{global} recruitment and defection rates.  Furthermore,  these global continuum descriptions of the underlying individual-level agent-based mechanisms faithfully capture the average behaviour of these agent-based simulations, providing not only more tractable and understandable mathematical models of sports riots, but also the crucial links between individual-level mechanisms and population-level phenomena.

There are several avenues for further consideration that stem from the modelling frameworks presented here. For instance, the ABM domain can easily be extended to incorporate additional  realistic features of a city layout, including a road and sidewalk network, public transport lines, and buildings. These additional movement augmentations and hindrances will clearly affect the direction and spread of riotous activity within the city structure.  Additionally, the incorporation of additional agent sub-populations, such as rival sports fans that are independently rioting,  would provide additional insight into the multifaceted nature of sports riots, such as to the relative effects between property damage and violent activity from opposing fans.
 Another feature that can be included in this ABM framework is the destructive nature of the rioters themselves. In this work, we simply consider the location and population density of the rioter sub-population, rather than what the rioters themselves are \textit{doing}. It would be beneficial to the application of sports riots, both from a mathematical and social sciences perspective, to incorporate `targets' of riotous activity, such as rival sports fans or nearby buildings and businesses. 
 Finally, the expansion of agent-based models into social science applications need not be contained to sports riots alone. For example, the worldwide phenomena of panic-buying amidst the COVID-19 pandemic also crucially hinges on what proportion of shoppers influence the recruitment or defection of panic-buying activity \cite{billore2021panic}.  The agent-based modelling framework presented in this work is an ideal starting point in terms of incorporating further aspects characteristic of panic-buying, such as dispersion and aggregation of shoppers \cite{starke2014nonlinear, d2006self}.  We leave these ABM extensions for future exploration.

\subsection*{Data accessibility}
All data and MATLAB algorithms used to generate results are available on Github at \url{https://github.com/nfadai/Clements2021}.

\bibliographystyle{vancouver}
\bibliography{RiotsProjectBib.bib}

\end{document}